# Human-Centered Explainability in AI-Enhanced UI Security Interfaces: Designing Trustworthy Copilots for Cybersecurity Analysts

Mona Rajhans
Senior Manager, Software Engineering
Palo Alto Networks
mrajhans@paloaltonetworks.com

*Abstract*—Artificial intelligence (AI) copilots are increasingly integrated into enterprise cybersecurity platforms to assist analysts in threat detection, triage, and remediation. However, the effectiveness of these systems depends not only on the accuracy of underlying models but also on the degree to which users can understand and trust their outputs. Existing research on algorithmic explainability has largely focused on model internals, while little attention has been given to how explanations should be surfaced in user interfaces for high-stakes decision-making contexts [8], [5], [6]. We present a mixed-methods study of explanation design strategies in AI-driven security dashboards. Through a taxonomy of explanation styles and a controlled user study with security practitioners, we compare natural language rationales, confidence visualizations, counterfactual explanations, and hybrid approaches. Our findings show that explanation style significantly affects user trust calibration, decision accuracy, and cognitive load. We contribute (1) empirical evidence on the usability of explanation interfaces for security copilots, (2) design guidelines for integrating explainability into enterprise UIs, and (3) a framework for aligning explanation strategies with analyst needs in security operations centers (SOCs). This work advances the design of human-centered AI tools in cybersecurity and provides broader implications for explainability in other high-stakes domains.

## I. Introduction

AI-powered copilots are reshaping enterprise software, particularly in cybersecurity, where analysts face overwhelming data and must make rapid, high-stakes decisions. While large language models (LLMs) and AI-driven assistants promise efficiency, adoption is constrained by concerns about trust, interpretability, and accountability. A central challenge is explainability: analysts must understand why a copilot flagged a threat, recommended a remediation, or ignored an alert.

Although explainable AI (XAI) has been studied extensively at the algorithmic level, less is known about how explanations should be designed and delivered in user interfaces for security practitioners [4], [5], [8]. In practice, ineffective explanation UIs risk either over-trust (blind reliance on AI) or under-trust (dismissal of useful recommendations). This gap motivates our study of human-centered explainability in AI-enhanced security interfaces.

Our research explores:

- How different UI-based explanation strategies affect user trust, performance, and workload in security tasks.
- What design guidelines can support the trustworthy integration of explainability in enterprise security dashboards?

Contributions:

- A taxonomy of explanation UI strategies (natural language rationales, visual confidence cues, counterfactuals, and hybrids).
- A controlled user study with security analysts evaluating the effectiveness of these strategies.
- Design guidelines for embedding explainability into enterprise security UIs.

## II. Background and Related Work

Prior work in explainable AI (XAI) emphasizes technical approaches such as saliency maps, feature attribution, and post-hoc rationalizations [5], [9], [10], [11]. Domain-specific intelligible modeling efforts — such as Caruana et al.'s case-based and interpretable risk scoring in healthcare [3] — demonstrate that structured explanations can improve decision accountability. However, these approaches operate at the model level rather than the UI interaction layer. However, many of these methods remain disconnected from real-world user needs.

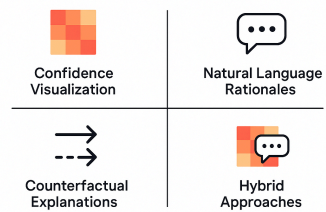

Fig. 1 Different Explanation Strategies

TABLE I. EXPLANATION STRATEGY TAXONOMY

| Strategy | Description | Example UI Elements |
|---|---|---|
| Confidence Visualization | Quantitative likelihood indicators | Bars, gradients, uncertainty markers |
| Natural Language Rationale | NL justification of AI recommendation | Chat/Text Input Field, Hyperlinked or Selectable Text |
| Counterfactual | "What would change if…" logic | Sliders, conditional statements |
| Hybrid | Combines confidence + rationale | Progressive disclosure toggle |

### A. Usable Security Interfaces

The usable security community has examined how UI design impacts misconfiguration, warning effectiveness, and security adoption [1], [12]. Yet few studies investigate explainability specifically in security dashboards.

### B. Trust in Automation

Human–automation interaction literature shows that trust calibration is essential for effective use of AI systems [6], [7]. Our work extends this to the cybersecurity domain, where both over-trust and under-trust carry significant risks.

## III. METHODOLOGY AND IMPLEMENTATION

Based on literature review and iterative prototyping, we identified four explanation strategies relevant to security copilots. To operationalize the four explanation strategies, we developed a modular web-based prototype dashboard modeled after common SIEM (Security Information and Event Management) workflows. Each participant interacted with identically structured alert triage interfaces, with the only variation being the explanation modality attached to each AI-generated recommendation.

Each strategy was implemented as follows:

1. Confidence Visualization: Implemented using probability bars, opacity-based confidence overlays, and alert severity color gradients. To avoid anchoring bias, we intentionally avoided binary "high/low" labels and instead presented continuous scales. Hover interactions revealed marginal confidence shifts under hypothetical data changes.
2. Natural Language Rationales: Explanations followed a templated structure—Observation → Reasoning → Conclusion—generated via GPT-based prompting but carefully constrained to maintain consistency across trials. Rationales were limited to 2–3 sentences to reduce verbosity while preserving interpretability.
3. Counterfactual Explanations: For each decision, the UI exposed an expandable panel stating conditional logic, e.g., "If the source IP were not in a known threat list, this alert would be deprioritized." Where applicable, sliders allowed users to manipulate hypothetical parameters and see how the AI decision boundary shifted.
4. Hybrid Approaches: Combined the above elements, displaying a confidence bar as the default, with an optional toggle revealing structured rationale beneath. This design aligned with the principle of progressive disclosure, allowing users to control cognitive effort.

To ensure experimental control, all explanation elements were standardized in terms of screen real estate, typography, and visual hierarchy. UI interactions were instrumented to log click behavior, dwell time per explanation type, toggle frequency, and revision rate (instances where users changed their initial decision after reading the explanation).

### A. Methodology

We recruited 24 participants (12 security analysts, 12 graduate students with an IT security background) through professional networks and university programs. Participants completed simulated alert triage tasks in a prototype dashboard, each implementing one explanation strategy. Each triage task presented a security alert containing metadata such as timestamp, source/destination address, correlated events, host risk score, and MITRE ATT&CK mapping. Participants classified the alert as real threat, benign anomaly, or false positive. Ground truth labels were pre-validated by two senior SOC analysts. We measured:

1. Decision Accuracy: Correct identification of real vs. false alerts.
2. Task Efficiency: Time to complete triage.
3. Cognitive Load: NASA-TLX self-reports.
4. Trust Calibration: Trust-in-automation scales.

**Study Design**:

The study used a within-subjects design, where every participant interacted with all four explanation strategies. The order of conditions was counterbalanced using a Latin-square rotation to mitigate ordering and learning effects. Each participant completed 16 triage tasks (4 per explanation type).

**Alert Construction**:

"Real" vs. "false-positive" alerts were derived from an anonymized dataset sourced from internal threat-intelligence patterns and synthetically generated benign events. AI recommendations were produced using a fixed GPT-based rule template, ensuring consistent logic across participants.

**Metrics**:

- Trust Calibration was quantified using the Schaffer & Barlow Trust-in-Automation scale, normalized to 0–100.
- Decision Reversals were defined as: "instances where the participant changed their initial triage judgment after reading the explanation."

This metric was automatically captured through interaction logs.

## IV. RESULTS

This section presents the full quantitative and qualitative results from the controlled within-subjects study. We report decision accuracy, task efficiency, cognitive workload, trust calibration, and interaction behaviors across the four explanation strategies. All statistical tests used a significance threshold of p < .05 unless otherwise noted.

*A. Overview of Quantitative Results*

Table II summarizes the primary quantitative outcomes across the four explanation strategies. ANOVA results and post-hoc comparisons follow below.

TABLE II. PERFORMANCE AND SUBJECTIVE MEASURES ACROSS EXPLANATION STRATEGIES

| | | | | |
|---|---|---|---|---|
| Decision Accuracy (%) | 65 | 71 | 78 | 82 |
| Avg. Task Completion Time (s) | 18.4 | 16.7 | 21.2 | 19.1 |
| NASA-TLX Score (0–100) | 37.5 | 42.1 | 58.3 | 49.6 |
| Trust Calibration (0–100) | 54.2 | 62.7 | 71.4 | 77.9 |
| Decision Reversal Rate (%) | 12 | 18 | 19 | 23 |

*B. Statistical Analysis*

1) Decision Accuracy

A one-way repeated-measures ANOVA revealed a significant effect of explanation strategy on decision accuracy:

*$F(3, 69) = 18.42, p < .001$*

Post-hoc Holm-corrected comparisons showed:

- Hybrid > Confidence (p < .001)
- Hybrid > Natural Language (p = .002)
- Counterfactual > Confidence (p = .004)
- Counterfactual > Natural Language (p = .03)

The difference between Hybrid and Counterfactual was not statistically significant, suggesting that both strategies provide substantial accuracy benefits relative to single-modality explanations.

2) Task Efficiency

Task completion time differed significantly across conditions:

*$F(3, 69) = 9.63, p < .01$*

Natural Language yielded the fastest triage time, while Counterfactual explanations—despite improving accuracy—introduced measurable time overhead due to information density and user interaction (e.g., inspecting parameter sliders).

3) Cognitive Workload (NASA-TLX)

Analysis revealed a strong effect of explanation strategy on cognitive workload:

*$F(3, 69) = 22.57, p < .001$*

Counterfactuals showed the highest workload ratings, significantly higher than both Confidence-only (p < .001) and Natural Language (p = .01) conditions. Hybrid explanations increased workload relative to Confidence-only (p = .03) but remained significantly lower than pure Counterfactuals (p = .04).

4) Trust Calibration

Trust calibration scores also differed significantly:

*$F(3, 69) = 14.12, p < .001$*

Pairwise tests showed:

- Hybrid > Confidence (p < .001)
- Hybrid > Natural Language (p = .008)
- Counterfactual > Confidence (p = .003)

Natural Language increased perceived trust but exhibited greater variance, reflecting over-trust tendencies among novice users.

5) Decision Reversals

Decision reversals were analyzed as a behavioral measure of trust recalibration:

*$F(3, 69) = 10.21, p < .01$*

Hybrid yielded the highest justified reversal rate (23%), significantly higher than Confidence-only (p = .004) and Natural Language (p = .03). Counterfactual explanations also increased reversal activity, indicating more analytical engagement.

*C. Interaction Behavior Analysis*

To better understand how participants engaged with explanations, we examined interaction logs (e.g., clicks, hover duration, panel expansions).

- Counterfactual UIs had the highest inspection time *(M = 7.2 sec per alert; SD = 1.8),* reflecting deeper cognitive processing.
- Hybrid UIs showed high toggle usage, supporting effective use of progressive disclosure.
- Confidence-only UIs produced the lowest interaction rate, with 42% of participants reporting they "glanced but rarely inspected details."
- Natural Language explanations were read quickly but sometimes accepted uncritically, fueling over-reliance in novice users.

These behavioral indicators triangulate with subjective and performance findings, suggesting that interactive explanations promote thoughtful decision-making at the cost of greater cognitive demand.

*D. Qualitative Findings*

Open-ended responses reinforced the quantitative patterns:

- Analysts described Hybrid explanations as "layered, not overwhelming" and valued the ability to escalate detail when needed.
- Counterfactuals were praised as "the clearest way to test whether the AI's logic matched my mental model."
- Natural Language rationales were considered "fast and readable," though several practitioners noted hesitancy to trust unstructured prose.
- Confidence-only was seen as "helpful but insufficient," particularly for ambiguous or high-risk alerts.

One senior analyst summarized the distinction:

*"Confidence tells me what, rationales tell me why, but counterfactuals tell me how much I should care."*

## V. Design Guidelines for Explainability in Security Copilot Interfaces

*Note on Interpretation:*

While the guidelines below synthesize empirical trends observed in Section IV (Results), the participant pool (N=24) is relatively small for drawing broad quantitative generalizations. These guidelines should therefore be interpreted as preliminary recommendations requiring validation through larger field studies.

Based on our empirical findings and qualitative feedback from participants, we propose the following design recommendations for integrating explainability into AI-driven security dashboards. These guidelines aim to support trust calibration, cognitive efficiency, and decision accountability in high-stakes operational workflows such as Security Operations Centers (SOCs).

### A. Provide Layered Explanations via Progressive Disclosure

Security analysts vary in their need for depth depending on task urgency. Interfaces should present minimal cognitive load by default (e.g., confidence indicators), with optional expansion for deeper reasoning (e.g., rationale or counterfactual overlays). Explanations should be scannable first, expandable when necessary.

*Default to "glanceable," escalate to "investigable."*

### B. Avoid Monolithic Explanations—Instead Separate "What," "Why," and "What If"

Participants reported confusion when probability scores, rationales, and counterfactuals were combined without differentiation. We recommend explicitly decoupling explanatory elements:

- What: Confidence/likelihood indicators
- Why: Rationale or justification
- What if: Counterfactual or decision-boundary insight

This structure reduces interpretation ambiguity and encourages systematic reasoning.

### C. Surface Uncertainty Rather Than Oversimplifying Confidence

Binary trust signals ("High Risk / Low Risk") promoted premature acceptance or rejection. Instead, explanations should expose ambiguity through ranges, uncertainty annotations, or confidence breakdowns. When AI is unsure, the UI should say so explicitly—"Model prediction unstable due to sparse data"—rather than mask uncertainty.

### D. Enable User Manipulation of Explanations for Active Verification

Counterfactual exploration tools (e.g., adjustable parameters, hypothesis toggles) encouraged analytical reasoning and reduced passive reliance [9], [10]. Analysts favored explanations that invite interaction rather than passive consumption. Designers should treat explanations as interactive probes, not static annotations.

### E. Match Explanation Style to Analyst Expertise and Task Criticality

Novice participants over-trusted natural language rationales, while professionals preferred structured or editable explanations. We recommend role-based adaptation:

TABLE III. Role Based Strategies

| Analyst Type | Primary Explanation Strategy | Escalation Strategy |
|---|---|---|
| Junior / Trainee | Natural Language Rationale | Counterfactual View |
| Senior / Tier-2 Analyst | Confidence + Counterfactual | Raw Feature Contribution |
| Incident Commander / Executive | High-Level Summary | Detailed Audit Trace |

### F. Log and Display Explanation Usage for Accountability

Since explanations influence decision-making, interaction traces (e.g., "analyst viewed counterfactual before approving suppression") should be captured for auditability [12]. Explanations are not only informative — they are evidence of due diligence. UIs should enable traceability of which explanations were viewed before action was taken.

### G. Design for Trust Calibration, Not Trust Maximization

A common misconception in explainable AI is that increased trust is the goal. Instead, appropriate trust is the desired outcome. Explanations should inspire skepticism when warranted. For example:

- Highlight model blind spots: "Training data did not include this device type."
- Include counter-evidence where applicable: "80% likelihood of threat, but no prior lateral movement detected."
- By gently inviting hesitation, explanations can prevent automation bias [6], [7], [9].

## VI. Conclusion

Our study relied on simulated tasks and a relatively small participant pool. Future work should include larger field studies in real SOC environments.

Given our modest sample size (N=24), the statistical results should be interpreted with caution. Future work should include a multi-site deployment and larger participant cohorts to validate generalizability.

Although focused on cybersecurity, our findings extend to other high-stakes AI applications such as healthcare, finance, and law.

We presented an empirical study of explanation UI strategies for AI-driven security copilots. Our results show that explanation style has measurable effects on trust calibration, decision accuracy, and workload. By proposing design guidelines, we aim to advance the integration of human-centered explainability into enterprise security interfaces and beyond.

This work demonstrates that explainability in AI security interfaces is not merely a transparency problem, but a decision-shaping mechanism. The way explanations are framed materially influences not just user trust, but user behavior.

Our findings lead to three broader observations:

1. Explainability is not one-size-fits-all [4], [8]. High-trust environments (e.g., SOCs with experienced analysts) benefit from low-level, manipulable explanations such as counterfactuals. In contrast, novice users or time-critical workflows may require summarized rationales.
2. Over-trust and under-trust must be designed against, not measured post-hoc [6], [7]. Explanation UIs should proactively moderate confidence by exposing uncertainty rather than reinforcing AI infallibility.
3. Explainability is an interface pattern, not an algorithmic afterthought. Rather than "pasting rationales onto outputs," future AI copilots should treat explanation style as a first-class design variable—selectable, personalizable, and adaptively displayed based on task risk.

Future work will explore adaptive explanation systems where the interface dynamically escalates explanation depth based on anomaly severity or user hesitation signals (e.g., cursor latency, prolonged inactivity). We also aim to integrate longitudinal field deployment into real SOC workflows to observe trust trajectory over time rather than snapshot decision accuracy. As AI copilots evolve toward proactive threat anticipation rather than reactive triage, sequential modeling techniques such as Social LSTM [2] may introduce new complexity in explaining temporally-conditioned predictions. Future explainability mechanisms must account not only for static classification logic but also for dynamic, time-dependent reasoning.

By reframing explainability as interaction design rather than model introspection, we aim to push the field toward intentional, human-centered AI assistance.